\documentclass[prb,twocolumn,showpacs,citeautoscript]{revtex4}
\usepackage{amsmath}
\usepackage{graphicx}

\newcommand{\degree}{\mbox{$^\circ$}}

\newcommand{\LTO}{LaTiO$_{3.41}$}
\newcommand{\LTo}{LaTiO$_{3}$}
\newcommand{\ABo}{\textit{AB}O$_{3}$}

\begin{document}

\preprint{\today}

\title{\boldmath Crystal structure of LaTiO$_{3.41}$ under pressure}

\author{I. Loa}
\email[Corresponding author:~E-mail~]{I.Loa@fkf.mpg.de}
\author{K. Syassen}
\author{X. Wang}
\affiliation{Max-Planck-Institut f{\"u}r Festk{\"o}rperforschung,
Heisenbergstra{\ss}e 1, D-70569 Stuttgart, Germany}

\author{F. Lichtenberg}
\affiliation{Institut f{\"u}r Physik, EKM, Experimentalphysik VI, Universit{\"a}t
Augsburg, Universit{\"a}tsstra{\ss}e 1, D-86135 Augsburg, Germany}

\author{M. Hanf\/land}
\affiliation{European Synchrotron Radiation Facility,
        BP 220, F-38043 Grenoble, France}

\author{C. A. Kuntscher}
\affiliation{Universit{\"a}t Stuttgart, 1.~Physikalisches Institut,
Pfaffenwaldring 57, D-70550 Stuttgart, Germany}

\date{\today}

\begin{abstract}
The crystal structure of the layered, perovskite-related \LTO\
(La$_{5}$Ti$_{5}$O$_{17+\delta}$) has been studied by synchrotron
powder x-ray diffraction under hydrostatic pressure up to 27~GPa
($T=295$~K). The ambient-pressure phase was found to remain stable up
to 18~GPa. A sluggish, but reversible phase transition occurs in the
range 18--24~GPa. The structural changes of the low-pressure phase are
characterized by a pronounced anisotropy in the axis compressibilities,
which are at a ratio of approximately $1:2:3$ for the $a$, $b$, and $c$
axes. Possible effects of pressure on the electronic properties of
\LTO\ are discussed.
\end{abstract}

\bigskip



\maketitle

\section{Introduction}

\begin{figure}
     \centering
     \includegraphics[width=0.96\hsize,clip]{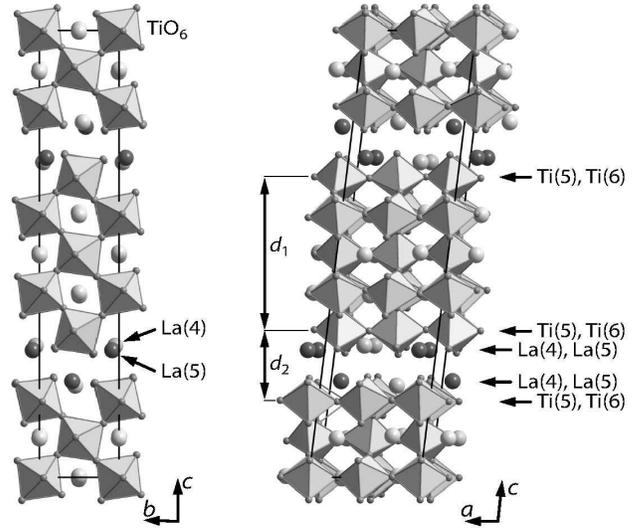}
     \caption{The monoclinic crystal structure \cite{DLS03} of \LTO\ at
         ambient conditions (space group $P2_1/c$, $Z=4$)  can be represented
         by slabs of corner-sharing TiO$_{6}$ octahedra, separated by layers
         of additional oxygen atoms. Within the slabs of thickness $d_1$, the
         octahedra are connected and tilted cooperatively in the same fashion
         as in the more familiar GeFeO$_3$-type distorted perovskites such as
         \LTo. La ions are located in the voids between the octahedra.}
     \label{fig:xtal}
\end{figure}

\LTO\ belongs to a series of perovskite-related layered
compounds of the composition
$A_nB_n$O$_{3n+2}$.\cite{DLS03,LHWM01,GC74,LWBW91}
\LTO\ is a $n=5$ member of this class, and thus it may alternatively be
denoted as La$_{5}$Ti$_{5}$O$_{17+\delta}$. Its monoclinic crystal structure
is illustrated in Fig.~\ref{fig:xtal}. The samples studied here and in
previous work \cite{LHWM01,KMDL03,DLS03} have a slight oxygen excess of 0.3\%
and are therefore denoted as \LTO\ instead of LaTiO$_{3.40}$.

As a consequence of the oxygen-rich composition of \LTO\ (compared to \LTo),
its Ti-$3d$ derived electronic bands are only partially occupied (Ti
$3d^{0.18}$). Electrical transport measurements\cite{LHWM01,KMDL03} on \LTO\
revealed strongly anisotropic properties and a quasi-1D metallic behavior. The
DC resistivity along the $b$ direction and perpendicular to the $ab$ planes
showed semiconducting behavior in the temperature range of 4--290~K. Along the
$a$ direction a metal-like temperature dependence of the conductivity was
reported for the temperature range of 60--200~K. Above 200~K a slight decrease
in resistivity with increasing temperature was observed, indicating conduction
due to polaron hopping\cite{LHWM01,KMDL03}. Below 60~K the resistivity
increased steeply, corresponding to an electronic activation energy of
$\sim$8~meV. The apparent opening of an electronic band gap has been discussed
in terms of a possible Peierls distortion of this quasi-1D
system\cite{LHWM01}.

A study of the mid- and far-infrared optical properties of
\LTO\ has recently corroborated its quasi-1D metallic character\cite{KMDL03}.
Furthermore, signatures of polaronic charge carriers were found in the form of
a mid-infrared band with pronounced temperature-dependent changes. The polaron
scenario is in overall agreement with the experimental results obtained so
far. The exact nature of the polaronic carriers, however, still needs to be
clarified.

The application of high hydrostatic pressure provides a tool to tune
the structural and the electronic properties of \LTO. On the one hand,
this may offer a possibility to drive the system more into a quasi-1D
metallic regime, i.e.\ over a wider range of temperatures. On the other
hand, and more importantly, it allows to investigate the nature of the
charge carriers, in particular their polaronic character, as was
detailed by Goddat~\textit{et~al.}\cite{GPP99}

Here we explore the structural changes of \LTO\ under hydrostatic pressures up
to 27~GPa. The crucial question of the structural stability of the
ambient-pressure phase is addressed. The pressure-induced structural changes
of the low-pressure phase are analyzed and the anticipated changes in the
electronic system discussed. We aim at providing a basis for the
interpretation of future investigations of the physical properties of \LTO\ at
high pressures.

\section{Experimental Details}

The structural properties of \LTO\  under pressure were studied up to 27~GPa
and ambient temperature by monochromatic ($\lambda = 0.4176$~{\AA}) x-ray powder
diffraction at the European Synchrotron Radiation Facility (ESRF Grenoble,
beamline ID9A). A crystal was ground finely and some powder placed into a
diamond anvil cell (DAC) for pressure generation. Nitrogen served as a
pressure transmitting medium to provide nearly hydrostatic conditions.
Diffraction patterns were recorded with an image plate detector and then
integrated \cite{soft:fit2d} to yield intensity vs.\ $2\theta$\/ diagrams. The
DAC was rotated by $\pm 3$\degree\ during the exposure to improve the powder
averaging. Pressures were determined with the ruby luminescence method
\cite{MXB86}. The diffraction diagrams were analyzed with the Rietveld method,
i.e.\ whole pattern fitting, using the GSAS software
\cite{soft:GSAS,GSAS-details}. The synthesis of the
\LTO\ material has been described elsewhere\cite{LHWM01}.

\section{Results and Discussion}

\begin{figure}
     \centering
     \includegraphics[width=0.9\hsize]{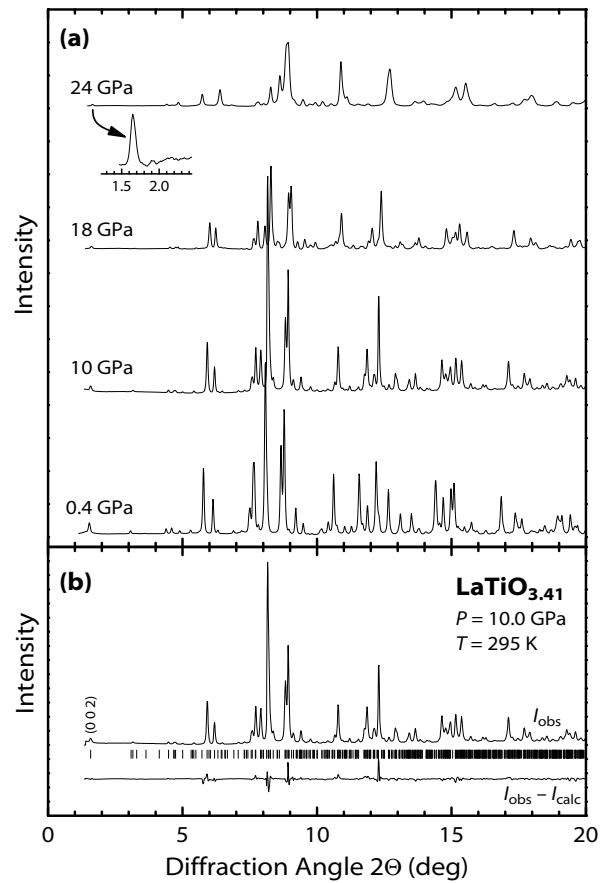}
     \caption{(a) Diffraction diagrams of \LTO\ at high pressures ($\lambda =
     0.4176$~{\AA}). The diffractogram of the high-pressure phase at 24~GPa shows
     a line at 1.65\degree\ (inset), corresponding to the $(002)$ reflection
     of the ambient-pressure phase.  (b) X-ray powder diffraction diagram
     ($I_\text{obs}$) of \LTO\ at 10.0~GPa and difference curve
     ($I_{\text{obs}}-I_{\text{calc}}$) for a partial Rietveld refinement (La
     and Ti positions only). Markers show the calculated peak positions.}
     \label{fig:diffpat}
\end{figure}

Figure~\ref{fig:diffpat}(a) shows x-ray diffraction diagrams of \LTO\ for
increasing pressures up to 24~GPa. At pressures above 2~GPa additional
reflections (mostly weak) due to various phases of solid nitrogen were
observed. Up to 18~GPa, there are no discontinuous changes in the diffraction
patterns of \LTO.

At pressures above 18~GPa a structural phase transition is evidenced by the
appearance of additional reflections. The transition is completed at 24~GPa.
There is a continuous evolution of the low-angle $(002)$ reflection of the
low-pressure phase towards a corresponding peak in the high-pressure diagrams
[inset of Fig.~\ref{fig:diffpat}(a)]. It suggests that the long $c$ axis
($c_0=31.5$~{\AA}) is preserved across the phase transition. As a result, there
are a large number of overlapping reflections, so that a unique determination
of the high-pressure unit cell has not been possible. The structural phase
transition is reversible.

A rather small increase in the widths of the reflections ($\sim$6\%) before
and after the $0 \rightarrow 27 \rightarrow 0$~GPa pressure cycle indicates
the creation of only a small amount of defects during the phase
transformations. It suggests that the phase transition may represent a
distortion of the low-pressure crystal structure rather than a reconstructive
transition.

\begin{figure}
     \centering
     \includegraphics[width=0.9\hsize]{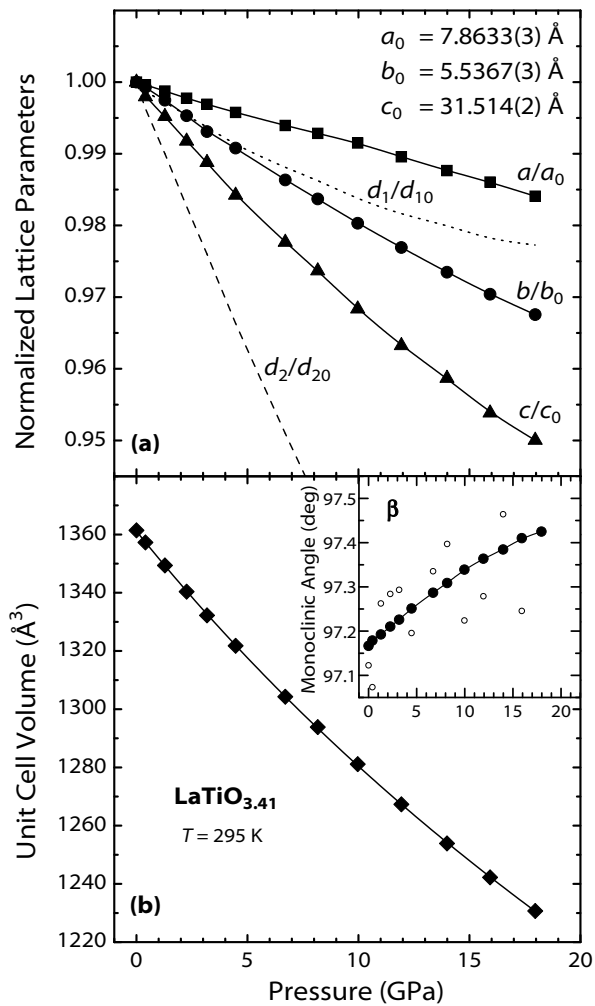}
     \caption{Structural parameters of \LTO\ as a function of pressure. (a)
     Lattice parameters $a, b, c$ as well as estimates of the slab thickness
     $d_1$ and interlayer separation $d_2$, normalized to their respective
     zero-pressure values. The zero-pressure values of $d_1$ and $d_2$ amount
     to $d_{10} \approx 11.14$~{\AA} and $d_{20} \approx 4.50$~{\AA}, respectively.
     See text for details. (b) The experimental pressure--volume data can be
     represented by a Murnaghan equation of state (solid line). The inset
     shows the variation of the monoclinic angle with pressure. Solid symbols
     represent angles determined from the lattice parameters $a$ and $c$; open
     symbols refer to the results of the refinements.
     }
     \label{fig:StructParam}
\end{figure}

The lattice parameters of the low-pressure phase as a function of pressure
were determined from Rietveld-type fits of the diffraction diagrams. This
approach was chosen over a direct determination on the basis of the  peak
positions because of the massive overlap of reflections
[Fig.~\ref{fig:diffpat}(b)]. The lattice parameters were determined up to
18~GPa as shown in Fig.~\ref{fig:StructParam}(a). The compression under
hydrostatic pressure is rather anisotropic. The relative compressibilities of
the $a$, $b$, and $c$ directions are at a ratio of approximately $1:2:3$.

The inset of Fig.~\ref{fig:StructParam}(b) shows that the monoclinic angle
increases slightly (from 97.17 to 97.43) with increasing pressure to 18~GPa.
It can be determined in two ways: either directly from the refinements (i.~e.,
from the peak positions) or from the ratio of the lattice parameters $a$ and
$c$. The latter is possible because there are two identical building blocks
per unit cell that are shifted by $a/4$ along the $a$ direction with respect
to each other [Fig.~\ref{fig:xtal}]. As a consequence, the monoclinic angle
$\beta$ is fully determined by the lattice parameters $a$ and $c$ through
$\beta = \arccos(a/2c)$. The values of $\beta$ calculated in this way have
much smaller uncertainties than those derived directly from the Rietveld
refinements, see Fig.~\ref{fig:StructParam}(b).

From the lattice parameters we calculate the unit cell volume as a function of
pressure as shown in Fig.~\ref{fig:StructParam}(b). The experimental data are
well represented by the Murnaghan relation \cite{Mur44} $V(P) = V_0 [(B'/B_0)
P+1]^{-1/B'}$ with the bulk modulus $B_0 = 142.2(11)$~GPa and its derivative
$B' = 4.3(2)$ at zero pressure. Here the ambient-pressure unit cell volume was
kept fixed at the experimental value of $V_0=1361.5(2)$~{\AA}$^{3}$. The
ambient-pressure unit cell volume determined here is $\sim$0.4\% larger than
reported previously \cite{DLS03}.

In order to explore the origin of the anisotropic compressibility, it is
worthwhile to estimate, as a function of pressure, the thicknesses $d_1$ and
$d_2$ of the \LTo-type slabs and the separating layers, respectively
[Fig.~\ref{fig:xtal}]. A measure of these quantities is provided by the
$z$-coordinates of the Ti(5) and Ti(6) ions that are adjacent to the
oxygen-rich layers. Therefore, an attempt was made to determine the atomic
positions of the heavier atoms La and Ti by means of Rietveld refinements.
Here one should bear in mind that the unit cell of
\LTO\ contains 20 atoms each of La and Ti with 15 La atomic coordinates  and
12 Ti coordinates to be determined. This is at the borderline of what can be
extracted from high-pressure x-ray powder diffraction data.

\begin{figure}
     \centering
     \includegraphics[width=0.8\hsize]{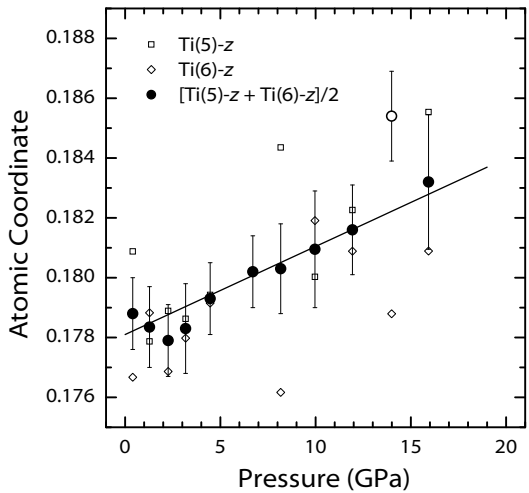}
     \caption{Evolution of the Ti(5) and Ti(6) $z$ coordinates as a
     function of pressure. Open squares and diamonds represent the Ti(5) and
     Ti(6) data, respectively, as determined in the Rietveld refinements.
     These data scatter in a correlated fashion. The circles show the averaged
     $z$ coordinates of Ti(5) and Ti(6). The solid line represents a linear
     fit to the data [$z(P) \approx 0.1781 + 0.00029 P/\text{GPa}$], with the
     outlier at 14~GPa (open circle) not being taken into account. The error
     bars mark three times the uncertainties reported by the refinement
     program GSAS.}
     \label{fig:Ti5+6pos}
\end{figure}

Figure~\ref{fig:Ti5+6pos} displays the variation with pressure of the Ti(5)
and Ti(6) $z$-coordinates. It is evident that the scatter in the values
determined directly from the refinements (smaller open symbols) originates
primarily from a correlation of larger Ti(5) with smaller Ti(6) $z$ positions.
At ambient conditions, the Ti(5) and Ti(6) ions are located at $z=0.1781 \pm
0.0006$, i.e., almost within a common plane. This is very similar to the \LTo\
case where the corresponding Ti atoms are placed strictly in a single plane.
Altogether, assuming equal $z$ parameters for the Ti(5) and Ti(6) atoms should
provide a good approximation to the actual situation. The averaged Ti(5,6) $z$
coordinates show a clear variation with pressure, see Fig.~\ref{fig:Ti5+6pos}.

From the Ti(5,6) $z$ coordinates we can estimate the thicknesses $d_1$ of the
\LTo-type slabs and $d_2$ of the separating layers [cf.\ Fig.~\ref{fig:xtal}].
Figure~\ref{fig:StructParam}(a) shows that the variation of $d_1$ with
pressure is comparable to that of the crystallographic $a$ and $b$ directions,
while the interlayer spacing $d_2$ is much more compressible. The large
compressibility along $c$ therefore results from the highly compressible
oxygen-rich layers where the rather strong Ti--O bonds are missing. This large
difference in compressibilities also explains the increase of the monoclinic
angle with pressure [inset of Fig.\ref{fig:StructParam}(b)].

As a check on the plausibility of the estimated $d_1, d_2$ values we have
calculated the bulk modulus of the \LTo-type layers; it amounts to $B_0
\approx 190$~GPa. While there seems to be no experimental data on the bulk
modulus of \LTo\ available for comparison, our estimate is close to the bulk
modulus $B_0 = 194(3)$~GPa of perovskite-type YTiO$_{3}$.\cite{WLS04:YTiO3}
Hence, the Ti(5,6) positions and the layer thicknesses derived therefrom can
be considered sufficiently accurate to gain some insight into the structural
changes within the unit cell.

\begin{figure}
     \centering
     \includegraphics[width=0.8\hsize]{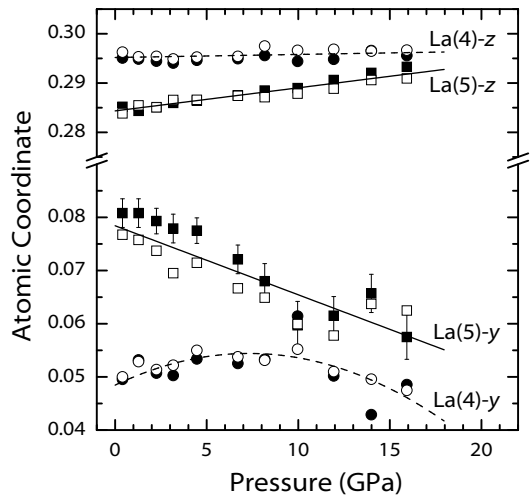}
     \caption{Evolution of the La(4) and La(5) atomic coordinates $y$ and $z$ as a
     function of pressure. Closed (open) symbols refer to refinement results
     where preferred orientation effects were (were not) taken into account.
     The error bars mark three times the uncertainties of the  La-$y$ coordinates reported
     by the refinement program GSAS. For clarity, they are shown only for one
     data set. In the case of the La-$z$ coordinates, the uncertainties are
     similar to the symbol size. Lines are guides to the eye.}
     \label{fig:La4+5pos}
\end{figure}

Additional information on the structural changes near the layer boundaries can
be obtained from the La(4) and La(5) positions. At ambient conditions both of
these ions are shifted from their regular positions ($z = 1/3$) between the
TiO$_{6}$ octahedra into the interlayer region [Fig.~\ref{fig:xtal}]. This
displacement is more pronounced for La(5) than for La(4).
Figure~\ref{fig:La4+5pos} illustrates that La(4) hardly moves along $c$ under
pressure, while La(5) shifts towards the \LTo-slab region. With regard to the
displacement off the $ac$ plane ($y=0$), La(5) moves continuously towards this
plane with increasing pressure while La(4) appears to be slightly more
displaced at intermediate pressures. Altogether, pressure forces the La(4) and
La(5) ions closer to the ``ideal'' positions in the $ac$ planes, they adopt
similar positions along the $c$ direction, but they remain displaced into the
interlayer region.

\begin{figure}
     \centering
     \includegraphics[width=0.7\hsize]{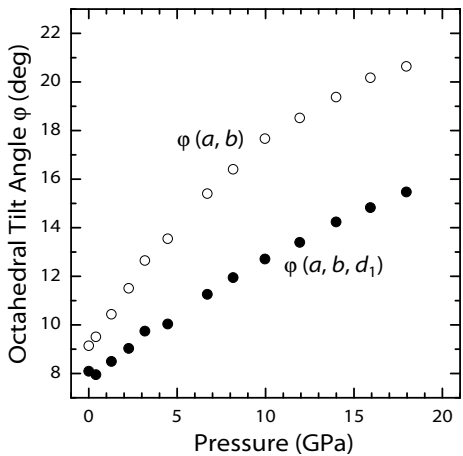}
     \caption{Octahedral tilt angles $\varphi$ \textit{vs} pressure derived from the
     lattice parameters $a$ and $b$ and the slab thickness
     $d_1$ according to Eqns (\ref{eq:1}) and (\ref{eq:2}).}
     \label{TiltAngles}
\end{figure}

With respect to the electronic properties of \LTO\ (and any other distorted
perovskite structure), the tilting of the TiO$_{6}$ octahedra represents a key
information. A measurement of these tiltings in the case of
\LTO\ would require an accurate determination of the 51 oxygen atomic
coordinates which does not appear feasible on the basis of an x-ray powder
diffraction experiment. It is possible, however, to arrive at a
semi-quantitative estimate of the tilt angles under pressure on the basis of
the lattice parameters and the slab thickness $d_1$. The relation between
octahedral tilting and orthorhombic distortion of the unit cell is well known
for the three-dimensionally linked perovskites of \ABo\ type
\cite{OH77,Gla72,MMLR95}. The same scheme can be applied to the \LTo-type
slabs of \LTO, where the octahedra are connected and tilted in the same
fashion. In the following, $d_1/2$ takes the role of the third lattice
parameter besides $a$ and $b$. On the basis of the equations given by O'Keeffe
and Hyde\cite{OH77}, we use the relations
%
%
\begin{align}
   \varphi(a,b,d_1) &= \arccos\left(\frac{\sqrt{8}\, b^2}{a d_1}\right) \label{eq:1}\\
\intertext{and}
   \varphi(a,b)     &= \arccos\left(6 (b/a)^2-2\right) \label{eq:2}
\end{align}
to determine the octahedral tilt angles shown in Fig.~\ref{TiltAngles}. The
geometrical constraints of the cooperative octahedral tilting that are at the
origin of these relations and the definition of $\varphi$ are detailed in
Ref.~\mbox{}\onlinecite{OH77}. Evaluation of the tilt angle on the basis of
Eq.~(\ref{eq:1}) with three lattice parameters is generally less sensitive to
distortions of the octahedra than the two-parameter determination. The
important result is that both approaches indicate a significant increase in
the average octahedral tilt angle $\varphi$ with increasing pressure. Roughly,
the tilting angle doubles at 18~GPa compared to ambient conditions.

The indication of a pressure-induced increase in octahedral tilt represents an
interesting difference in comparison to other transition metal perovskites
such as LaMnO$_{3}$ \cite{LAGS01,PRDG01}, PrNiO$_{3}$ \cite{MMLR95}, and
NdNiO$_{3}$ \cite{Amb03:LTO} which exhibit a reduction in the tilt. While
there are certainly no universal pressure-induced changes in the
GdFeO$_{3}$-type perovskite compounds, as has been pointed out
before\cite{MMLR95}, there may exist a systematic behavior across the series
of \textit{RT}O$_{3}$ rare-earth transition-metal oxides that appears
worthwhile to be explored.

Stability of \LTO\ under pressures of up to 18~GPa opens way to tune its
electrical transport properties. On the one hand, one can vary the itinerancy
of the system that is determined by the Ti--O bond lengths and Ti--O--Ti bond
angles. A reduction in bond lengths results in a larger bandwidth $W$ of the
electronic bands derived from the $t_{2g}$-orbitals and an enhanced itinerancy
of the system, while a decrease in the bond angles (increased octahedral tilt)
has the opposite effect. Hence, we have here two competing pressure-induced
changes that will affect the itinerancy of the system. These changes are most
important for the quasi-1D metallic state that exists in the temperature range
of 60--200~K at ambient pressure. On the other hand, polaron binding energies
are generally expected to decrease with increasing pressure due to the
stiffening of the lattice.\cite{GPP99} Therefore, high-pressure optical and/or
electrical transport experiments are promising approaches to test the
hypothesis on polaronic conductivity in the quasi-1D \LTO.

In conclusion, we have studied the crystal structure of \LTO\ by synchrotron
x-ray powder diffraction up to 27~GPa. The ambient-pressure phase remains
stable up to 18~GPa (at 295~K). Above 18~GPa a sluggish phase transition
occurs, which is completed at 24~GPa. The low-pressure phase is characterized
by a pronounced anisotropy of the axis compressibilities at a ratio of
approximately $1:2:3$ for the $a$, $b$ and $c$ axes. The anisotropy can be
related to rather compressible oxygen-rich layers that separate the
perovskite-type slabs and to variations in the tilt angles of the TiO$_{6}$
octahedra. Stability of \LTO\ up to 18~GPa opens way to tune its electronic
transport properties over a relatively large range. It motivates further
electrical transport and optical experiments to study the unusual transport
properties of \LTO\ in more detail.




\begin{thebibliography}{10}
\bibitem{GSAS-details}
\bibinfo{note}{Owing to the large number of adjustable atomic coordinates (15
  $\times$ La, 12 $\times$ Ti, 51 $\times$ O), only the positions of the
  heavier atoms La and Ti were refined. Usually, the analysis was performed in
  at least two ways, i.~e.\ without and with preferred orientation corrections
  applied. The differences in the results mark the experimental uncertainty.
  Preferred orientation effects were modeled using spherical harmonics. A
  simple pseudo-Voigt line shape was sufficient to describe the reflection
  profiles up to the highest pressures.}
\end{thebibliography}

\printtables
\printfigures

\end{document}